\documentclass[runningheads]{llncs}
\usepackage[hidelinks]{hyperref}
\usepackage{graphicx}

\usepackage[english]{babel}
\usepackage{amsmath}	%
\usepackage{microtype}
\usepackage{float}
\usepackage{paralist}
\usepackage{booktabs}
\usepackage{adjustbox}
\usepackage{amssymb}
\hypersetup{
  colorlinks   = true, %
  urlcolor     = NavyBlue, %
  linkcolor    = NavyBlue, %
  citecolor   = OliveGreen %
}
\usepackage[margin=0.15cm]{caption}
\usepackage{wrapfig}
\usepackage[dvipsnames]{xcolor}
\usepackage{xspace}

\DeclareMathOperator{\coss}{cosS}
\DeclareMathOperator{\actvec}{act2vec}

\newcommand{\mypar}[1]{\smallskip\noindent\textbf{#1.}}

\newcommand{\coselog}{\emph{CoSeLoG}\xspace}
\newcommand{\bpic}{\emph{BPIC 2020}\xspace}
\newcommand{\sepsis}{\emph{Sepsis}\xspace}

\begin{document}

\title{A Distance Measure for Privacy-preserving Process Mining based on 
Feature Learning}
\titlerunning{Distance Measure for Privacy in Process Mining based on Feature Learning}
\author{Fabian Rösel\inst{1}\and
	Stephan A. Fahrenkrog-Petersen \inst{1}\and
	\\Han van der Aa\inst{2} \and
	Matthias Weidlich\inst{1}}
\authorrunning{Fabian Rösel et al.}
\institute{Humboldt-Universität zu Berlin , Berlin, Germany \and
	University of Mannheim,  Mannheim, Germany\\
	\email{fabian.roesel@hu-berlin.de}, 
	\email{stephan.fahrenkrog-petersen@hu-berlin.de}, 
	\email{han@informatik.uni-mannheim.de},
	\email{matthias.weidlich@hu-berlin.de}}
\maketitle              %
\begin{abstract}
	To enable process analysis based on an event log without compromising the 
	privacy of individuals involved in process execution, a log may 
	be anonymized. 
Such anonymization strives to transform a log so
	that it satisfies provable 
	privacy guarantees, while largely maintaining its utility for process 
	analysis. 
	Existing techniques perform anonymization using simple, syntactic measures 
	to identify suitable 
	transformation operations.
	This way, the semantics of the activities referenced by the events in a 
	trace are neglected,
	potentially leading to transformations in which events of unrelated 
	activities are merged.
	To avoid this and incorporate the semantics of activities during 
	anonymization, 
	we propose to instead incorporate a distance measure 
	based on feature learning. 
	Specifically, we show how embeddings of events enable the 
	definition of a distance measure for traces to guide event log 
	anonymization. Our experiments with 
	real-world data indicate that anonymization using this 
	measure, compared to a syntactic one, yields logs that are closer 
	to the original log in various dimensions and, hence, have higher 
	utility for process analysis.
\end{abstract}
\section{Introduction}
\label{sec:introduction}

Privacy-preserving process mining~\cite{elkoumy2021privacy} aims at 
protecting personal data, while at the 
same time, enabling organizations to improve their business processes. 
The consideration of privacy in process mining is necessary, since the event 
logs used as a starting point for the analysis often include information about 
the individual people involved in process execution. 
For example, for a treatment process in a hospital, an event log may include 
personal information about the treated 
patients~\cite{stefanini2018performance}. Since the respective data was 
commonly not recorded for the purpose of operational improvement, process 
mining is often considered a secondary use of the data and, as such, strictly 
regulated.

In general, there are two angles to approach privacy-preserving process 
mining~\cite{fahrenkrog2019providing}: One may anonymize the 
event log used for process mining~\cite{pripel} or one may design the 
algorithms for process mining such that the output satisfies privacy 
guarantees~\cite{kabierski2021}.

An established strategy for anonymizing event logs is to group traces together, 
i.e., to merge the data of multiple executions of a process. This way, the 
information about a single process instance and, hence, about an individual 
person involved therein, is not revealed. Such an 
approach is realized, for instance, in PRETSA~\cite{PRETSA} and the 
TLKC framework~\cite{rafiei2020tlkc}. In order to preserve the utility of the 
event log for process analysis, these anonymization algorithms incorporate a 
distance measure. Intuitively, by merging traces that are close to each other, 
the resulting log will be close to the original one. Since process mining 
primarily targets the analysis of the behaviour exhibited by the traces, their 
closeness is captured in terms of their behavioural similarity. 

Yet, existing algorithms~\cite{PRETSA,rafiei2020tlkc} employ
simple syntactic measures, such as the Levenshtein distance.
 This is a 
limitation, since, contrary to the assumption behind the Levenshtein distance, 
not all events have the same closeness.
Arguably, the semantics of 
the activities referenced in the events in terms of their context suggests that 
some events are closer to each other than others.
 Neglecting such semantics means that any event log anonymization approach
may induce a higher loss in utility than 
what would be necessary to achieve a certain privacy guarantee. 

In this paper, we therefore investigate alternative approaches to measuring the 
distance of traces when anonymizing event 
logs. We study how embeddings of events, i.e., feature vectors 
that provide a semantic representation of the events, may be used as a 
foundation for a distance measure. Specifically, we rely on the 
Act2Vec~\cite{de2018act2vec} model to learn the embeddings from the original 
event log, thereby capturing the context of the activities of the events. 
Based thereon, we design a distance measure for traces that is tailored to 
event log anonymization, as it indicates which traces shall be merged into 
which other traces when aiming at a small loss of the log's utility for process 
analysis. Finally, we report on evaluation 
experiments to shed light on the impact of the adopted distance measure on the 
utility of an anonymized log. For several real-world event logs, we observe 
that using our semantic distance measure instead of a syntactic one, yields 
event logs that are closer to the original log along various dimensions.

In the remainder, we first discuss essential notions and notations 
(\autoref{sec:background}). We then present an embedding-based trace distance 
measure to use in event log anonymization 
(\autoref{sec:approach}). Finally, we report on evaluation experiments 
(\autoref{sec:evaluation}), before we review related work 
(\autoref{sec:related_work}) and conclude (\autoref{sec:conclusion}).

\section{Background}
\label{sec:background}

Below, we provide background for our work. In \autoref{sec:logs}, we 
introduce a model for event logs. \autoref{sec:pretsa} 
outlines the PRETSA algorithm, a state-of-the-art algorithm for event log 
anonymization. In \autoref{sec:act2vec}, we review the Act2Vec model to 
learn a feature-based representation of the events of a trace. 

\subsection{Event Logs}
\label{sec:logs}
Process Mining relies on event logs that capture the execution of business 
processes~\cite{van2012process}. We consider a common model of event logs, 
summarized as follows. 

Each step of a process is represented by an activity $a \in\mathcal{A}$, with 
$\mathcal{A}$ denoting the universe of all activities. Each execution of such 
an activity is represented as an event $e = \langle c, a, ts \rangle$, where 
$a$ is the respective activity, $ts$ is a timestamp specifying the time of 
activity execution, and $c$ is a case identifier signalling the instance of the 
process that the activity execution was related to. Moreover, by~$\mathcal{E}$, 
we denote the universe of all events.

All events with the same case identifier form a trace, an ordered sequence of 
events 
$\langle e_1, \ldots, e_n \rangle = t\in \mathcal{E}*$, with $t(i)\in 
\mathcal{E}$, $1\leq i \leq n$ denoting the $i$-th event of the trace. Here, 
the order of events in a trace is induced by their 
timestamps. We write $|t|$ for the trace length, i.e., the number of its 
events. A finite set of traces forms an event log, denoted by $L\subseteq 
\mathcal{E}*$. 

Three traces of an exemplary event log for a process to handle loan 
requests are shown in \autoref{tab:example_events}. After a request has 
been checked, the interest rate is negotiated or calculated automatically, a 
contract is set up, and the client is informed or gets the contract posted by 
mail. However, the check may also identify fraud, so that the account is 
blocked, before the client is informed about~it. 

\begin{table}[t]
	\caption{Three example traces of a request handling process.}
	\label{tab:example_events}
	\vspace{0.5em}
	\scriptsize
	\centering
	\begin{tabular}{l l r}
		\toprule
		Case & Activity & Time\\
		\midrule
		1 & Check Loan Req. & 9:05\\
		1 & Negotiate rate & 10:04\\
		1 & Set up contract & 10:45\\
		1 & Inform client & 14:08\\
		\bottomrule
	\end{tabular}\hspace{1em}
	\begin{tabular}{l l r}
	\toprule
	Case & Activity & Time\\
	\midrule
	2 & Check Loan Req. & 7:37\\
	2 & Calculate rate & 7:45\\
	2 & Set up contract & 8:25\\
	2 & Mail contract & 9:50\\
	\bottomrule
\end{tabular}\hspace{1em}
	\begin{tabular}{l l r}
	\toprule
	Case & Activity & Time\\
	\midrule
	3 & Check Loan Req. & 9:49\\
	3 & Report fraud & 10:12\\
	3 & Block account & 10:16\\
	3 & Inform client & 11:02\\
	\bottomrule
\end{tabular}
	\vspace{-1.5em}
\end{table}

\subsection{Event Log Anonymization with PRETSA}
\label{sec:pretsa}

Striving for privacy-preserving process mining, an event log may be anonymized 
to avoid disclosure of sensitive information. 
This approach is realized in the \textit{pre}fix-\textit{t}ree-based event log 
\textit{sa}nitization~\cite{PRETSA}, or PRETSA for short. The algorithm 
transforms a given log to provide privacy guarantees based on 
$k$-anonymity~\cite{kanonymity} and $t$-closeness~\cite{t-closeness}, 
while aiming to maximize the utility of the anonymized log.

The general idea behind PRETSA is protect the characteristics of a trace that 
enable the identification of an individual person. Specifically, it considers 
linkage attacks that are based on activity occurrences that directly point to 
individual persons, or exploit sensitive attributes for which differences in 
the observed distributions enable conclusions on individual persons.
PRETSA guards against the former attack, by adopting a notion of $k$-anonymity, 
i.e., it forms groups of traces that are undistinguishable from each other and 
comprise at least $k$ members. This way, the chances of linking a trace to an 
individual person is limited to ${1}/{k}$. To avoid linkage attacks through 
sensitive attributes, PRETSA also ensures $t$-closeness, i.e., it limits the 
difference in the distribution of a sensitive attribute of a group compared to 
the overall population.

To achieve the above guarantees, PRETSA transforms an event log by merging 
similar traces. It detects trace prefixes that violate the 
privacy guarantees and merges them into a similar variant. 
The latter is selected based on a closeness measure to limit the loss in 
utility caused by the 
transformation. However, PRETSA (and related approaches) assess the similarity 
of traces solely with a simple syntactic measure, i.e, the Levenshtein 
distance. Hence, all events of a trace are assumed to be 
equally close, which neglects the semantics of the respective activities, i.e., 
the context in which they are executed.

\subsection{The Act2Vec Model}
\label{sec:act2vec}

To incorporate the semantics of data elements, embeddings, i.e., feature 
vectors derived from a learned model, have been adopted in various domains. 
Most prominently, in natural language processing, an embedding can be 
constructed for each word of a text in order to represent its semantic 
meaning~\cite{word2vec}. 
A model to derive the embedding of a word is learned by 
considering the surrounding words and, hence, captures the context in which a 
word typically appears.

The idea of word embeddings was recently lifted to event logs, as part of 
the so-called Act2Vec model~\cite{de2018act2vec}. This way, one may learn 
semantic representations of activities and, therefore, events. 
As in the case of word embeddings, these representations are learned from the 
context of an activity: If an event indicates the execution of a specific 
activity, the surrounding events in traces of the event log indicate which 
activities represent the common execution context. Consequently, closeness 
of activities (and hence, events) is assessed based on the closeness of their 
execution context, 
rather than other properties, such as the labels.

\section{An Embedding-based Trace Distance Measure for Event Log Anonymization}
\label{sec:approach}

This section introduces a distance function for traces that incorporates the 
semantics of activities and, hence, promises to induce a lower loss of utility 
when used in event log anonymization. First, we discuss the intuition of the 
measure (\autoref{sec:intuition}), before we turn to its definition 
(\autoref{sec:measure}). 

\subsection{Intuition}
\label{sec:intuition}

Reconsider the three traces from \autoref{tab:example_events}. 
When comparing the first trace (Case 1) to the others, it is clear that Case 1 is more similar to the second trace than to the third: the first two both capture the regular handling of a loan request, differing  solely in the way the interest rate is determined (negotiated versus automatically calculated) and in the method used to get back to the client (informed versus per mail). 
The third trace (Case 3), in turn, denotes a very 
different scenario, in which fraud was detected, resulting in the account being blocked.

In terms of the purely syntactic Levenshtein distance, however, the second and 
the third trace are equally distant from the first one. Both of them include 
two activity executions that are in line with the first trace, whereas each of them also includes two events that are without counterpart in the first trace.
Therefore, existing anonymization techniques would consider both traces to be equally suitable in a merging step, even though merging the first and the third trace could be expected to  lower the utility of the resulting log more 
drastically than the alternative solution of merging the first and the second 
trace.

Instead, when the semantics of activities in terms of their execution context 
are incorporated, a more suitable assessment of the traces' similarity may be 
achieved.
For instance, the ``\emph{Negotiate rate}'' and ``\emph{Calculate rate}'' activities 
can be expected to have very similar execution 
contexts, i.e., after the request has been checked and directly preceding the 
activity to set up the contract. Hence, events referring to these activities 
are closer than events of less related activities, such as fraud reporting or 
account blocking.

\subsection{A Distance Measure for Traces based on Embeddings}
\label{sec:measure}

Feature learning based on the Act2Vec model~\cite{de2018act2vec} enables us to 
incorporate the semantics of the activities referenced in events in terms of 
their context of execution. The latter is induced by behavioural relations that 
describe common predecessors and successors of the respective activity. Since 
many process mining tasks are 
grounded in exactly these behavioural relations, preserving them as much as 
possible will increase a log's utility for process analysis.  

\smallskip
\noindent
\textbf{Event distance.}
Since embeddings encode the context of activities referenced in events, 
i.e., their predecessors and successors,
 the similarity (or the distance, respectively) of two traces can be derived 
 from the similarity (or distance) of their individual events.
Therefore, we first define a distance for events, before using it as the basis 
to quantify the distance of traces. 

We compare two events by the Cosine similarity of their feature vectors, which 
is defined over a vector space $V$ as: 
\begin{align*}
\coss: V \times V \rightarrow [-1,1],\quad \quad \quad 
(\textbf{v}{_1}, \textbf{v}{_2}) \mapsto \frac{\textbf{v}{_1} \cdot 
\textbf{v}{_2}}{||\textbf{v}{_1}||{_2} \: ||\textbf{v}{_1}||{_2}},
\end{align*}
with $||\textbf{v}||{_2}$ as the Euclidean norm of a vector $\textbf{v}$. Note 
that two vectors $\textbf{v}{_1}$, $\textbf{v}{_2}$ are identical in direction, 
if $\coss(\textbf{v}{_1}, \textbf{v}{_2}) = 1$; orthogonal, if 
$\coss(\textbf{v}{_1}, \textbf{v}{_2}) = 0$, and opposing, if
$\coss(\textbf{v}{_1}, \textbf{v}{_2}) = -1$.

In the remainder, given two events $e_1,e_2\in \mathcal{E}$, we capture their 
distance based on embeddings as $d_e(e_1,e_2) = 1 - 
\coss(\actvec(e_1),\actvec(e_2))$. 

\begin{figure}[t]	%
	\centering
	\includegraphics[clip,trim=0em 1.4em 0em 0em, 
	width=0.9\linewidth]{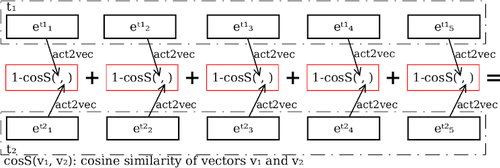}
	\vspace{-.5em}
	\caption{Visual representation of the measure, assuming $\vert t_1\vert = 
	\vert t_2\vert$.}
	\label{fig:measure}
	\vspace{-1.5em}
\end{figure}

\smallskip
\noindent
\textbf{On the direction of a trace distance for log 
anonymization.}
To assess the distance of two traces, the pair-wise distance of the events at 
the respective positions is calculated and summed up. Assuming that two traces 
$t_1$ and $t_2$ have equal length, this idea is illustrated in 
\autoref{fig:measure}. If two traces have a different length, the measure needs 
to take into account that, conceptually, events need to be added or removed to 
transform one trace into the other one. We therefore incorporate a penalty that 
is linear in the number of such excess events. 

However, to use 
the distance measure to decide which traces to merge during event log 
anonymization,  we define 
the penalty for excess events and, hence the trace distance measure, 
in an asymmetric manner.
Algorithms for event log anonymization such as 
PRETSA merge one trace into another, which is an inherently directed 
operation. Here, it is 
desirable to avoid introducing new behaviour as part of the anonymization in 
order to maintain the log's utility for process analysis. 
Practically, PRETSA does not add new trace 
variants, i.e., all sequences of activity executions of traces in the 
anonymized log are already represented by at least one trace in the original 
log. As such, the algorithm guarantees that the successor relation 
over activities as derived from traces may only be reduced by the 
anonymization. 
The above is achieved by removing excess events of a trace when merging it into 
another one, which implies a potential loss of information on the context in 
which activities are executed. 
Hence, 
it is preferred to merge shorter traces into 
longer ones in order to maintain a larger part of the successor relation. 

Moreover, the preservation of longer traces also helps to achieve privacy 
guarantees through the transformation. If traces of a certain variant, 
i.e., of a certain sequence of activity executions, are less than k-frequent, 
they have to be merged into other traces to achieve k-anonymity. Preferring the 
preservation of longer traces, there is a higher chance that these traces can 
be merged into traces that include the respective sequence of activity 
executions already as a prefix. In that case, no information in terms of the 
successor relation would be lost. 

\smallskip
\noindent
\textbf{Trace distance.}
Following the above arguments, we propose a directed 
distance measure for traces to guide event log anonymization. 
Given two traces $t{_1}$ and $t{_2}$, this distance incorporates that, if both 
traces have different lengths, merging the shorter into the longer one is 
preferred. The distance $d(t_1,t_2)$ is defined as:  
\[
d(t_1,t_2) = 
\begin{cases}
\sum\limits_{1\leq i \leq |t_1|} {d_e(t_1(i),t_2(i))} 
	& \text{if } |t_1| = |t_2|\\
\sum\limits_{1\leq i \leq |t_2|} 
\left( d_e(t_1(i),t_2(i)) \right)
+ (|t_1| - |t_2|) \cdot \rho_A & \text{if } |t_1| > |t_2|\\
\sum\limits_{1\leq i \leq |t_1|} 
\left( d_e(t_1(i),t_2(i)) \right)
+ (|t_2| - |t_1|) \cdot \rho_R & \text{if } |t_1| < |t_2|\\
	\end{cases}
	\]
where $\rho_A, \rho_R \in \mathbb{N}$ are penalties for event addition and 
event removal, respectively, such that $\rho_A>1$ and $\rho_R> \rho_A$.

As the distance measure incorporates the direction in which traces are 
potentially merged, the 
penalty $\rho_R$ for event removal shall be larger than the one for event 
addition, $\rho_A$. Empirically, we found that a 50\% increase of the penalty 
yielded the best results. Moreover, both penalties shall dominate the distance 
of any pair of individual events. Hence, we set $\rho_A=2$ and $\rho_R=3$ 
as default values.

\section{Evaluation}
\label{sec:evaluation}

In the following section, we present an evaluation of our approach. The aim of 
this evaluation is to answer the following research questions:
\begin{description}
	\item[RQ1:] Can we preserve more control-flow information by incorporating 
	an embedding-based trace distance measure into event log anonymization?
	\item[RQ2:] Do we preserve less utility in aspects unrelated to 
	control-flow by optimizing for higher control-flow related utility?
	\item[RQ3:] How do the internal properties of the distance measure impact 
	the preserved utility? 
\end{description}

To answer these questions, we apply our approach on the logs presented in 
\autoref{sec:dataset}, before \autoref{sec:exp_setup} lays out the experimental 
setup. \autoref{sec:results} presents and interprets the obtained results.

\subsection{Dataset}
\label{sec:dataset}

We perform the evaluation on three real-world event logs. As shown in 
\autoref{tab:dataset}, the logs differ considerably in their domains and 
structuredness. 
The \coselog log captures a relatively structured process, concerning the 
application of environmental permits, whereas the \sepsis log covers an unstructured process corresponding to clinical pathways from a hospital. 
Finally, \bpic log captures the 
semi-structured reimbursement process of travel costs at a Dutch university. 
While all three event logs have comparable size in terms of the number of 
cases, and hence traces, the differences between them in terms of 
structuredness are clearly illustrated by the average and maximum numbers of 
cases per trace variant. In \coselog, hundreds of traces may follow the same 
sequence of activity executions, whereas in \sepsis, traces often show a unique 
activity sequencing. 

\begin{table}[h!]
\vspace{-2em}
\caption{Event log characteristics.}
\label{tab:dataset}
\centering
\begin{tabular}{l @{\hspace{.8em}} r @{\hspace{.8em}} r @{\hspace{.8em}}r 
@{\hspace{.8em}}r}
\toprule
\textbf{Event log} & \textbf{Cases} & \textbf{Variants} & \textbf{avg. cases/var.} & \textbf{max. cases/var.}\\
\midrule
\coselog\cite{APLog} & 1,434 & 116 & 12.4 & 713 \\
 \sepsis\cite{SCLog} & 1,050 & 846 & 1.2 & 35 \\
 \bpic\cite{TravelLog} & 2,099 & 896 & 2.3 & 206 \\
\bottomrule
\end{tabular}
\vspace{-2em}
\end{table}

\subsection{Experimental Setup}
\label{sec:exp_setup}

\mypar{Configuration}
In our experiments, we compare the proposed method of incorporating  the embedding-based distance measure against the original PRETSA using Levenshtein distance.
To vary the required privacy guarantees, $k$-anonymity and $t$-closeness, we evaluate all combinations with $k = 2^{i}$ for $i \in [1, 8]$ and $t \in \{0.1, 0.25, 0.5, 0.75, 1.0\}$, yielding a total of 40 different settings per event log.

\mypar{Evaluation measures}
To asses the utility of anonymized event logs, 
we use two metrics to analyse the preservation of control-flow behaviour and one metric to measure the impact on the average cycle time of each activity:

\smallskip \noindent 
\emph{Advanced Behavioural Appropriateness.}
To measure the impact that anonymization has on the control-flow, we first use advanced behavioural appropriateness~\cite{rozinat2008}, which encodes behaviour within a log according to binary activity relations.
Specifically, given a log, an activity $A$ may \{\emph{always, sometimes, never}\} indirectly \{\emph{follow, precede}\} another activity $B$. 
By comparing these relations in an anonymized event log and its original counterpart, the metric yields a similarity score $\in$ [0,1], where 1 represents total equality for every relation between activity pairs in the two event logs and 0 if none of the relations are equal.

\smallskip \noindent 
\emph{Truly sampled Behaviour.}
We also use \emph{truly sampled behaviour}~\cite{trulySampled} as a metric to measure the preservation of control-flow information. While the behavioural appropriateness considers indirectly-follows relations between activities, this measures looks at directly-follows relations and also consider their relative frequency. 
Specifically, it quantifies what fraction of directly-follows relations of the original log are \emph{appropriately sampled}. 
It considers a directly-follows relation appropriately (or truly) sampled when, 
adjusted for potentially different log size, it appears approximately the same 
time in both logs. The \emph{Truly Sampled Score} then is the proportion of 
truly sampled directly-follows relations. Therefore a Truly Sampled Score of 
100\% is considered optimal, while a score of 0\% indicates no relation-wise 
similarity between the compared event logs.

\smallskip \noindent 
\emph{Total Duration Error.}
Finally, we also investigate how our approach impacts the duration of an activity, which can be regarded as a sensitive attribute that shall be anonymized in accordance with a specified $t$-closeness guarantee. 
For this, we compute the total duration error.
It is a simple metric based on the assumption that similar activities take a similar amount of time to execute. %
The error is calculated by comparing the total time it takes to execute all traces in the sanitized log to the total execution time of the original log. %

\mypar{Implementation}
To conduct our experiments, we implemented PRETSA using the embedding-based distance measure, as well as all used evaluation measures in Python. The source code is available at GitHub\footnote{\url{https://github.com/roeselfa/FeatureLearningBasedDistanceMetrics}} under MIT license.

\subsection{Results}
\label{sec:results}

\begin{figure}[!ht]
\begin{minipage}{0.5\textwidth}
\includegraphics[width=\linewidth]{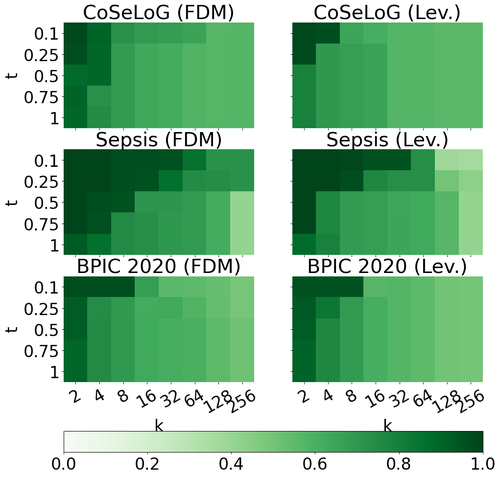}
\caption{Advanced Behavioural Appropriateness of anonymized logs.}
\label{fig:AdvancedBA}
\end{minipage}
\begin{minipage}{0.5\textwidth}
\includegraphics[width=\linewidth]{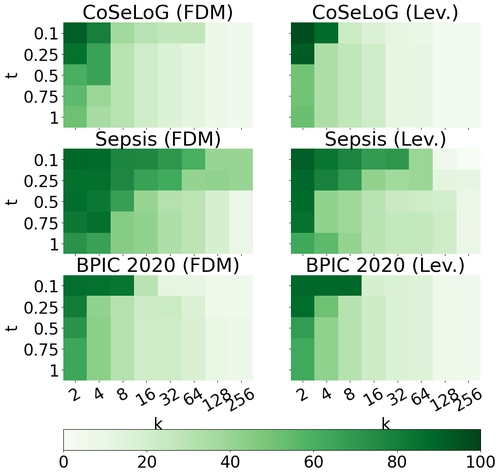}
\caption{Truly Sampled scores of anonymized logs.}
\label{fig:TSS}
\end{minipage}
	\vspace{-2em}
\end{figure}

To answer RQ1, we consider the evaluation metrics related to the control-flow 
behaviour of the log.  \autoref{fig:AdvancedBA} illustrates that our 
feature-based distance measure (FDM) can improve Advanced Behavioural 
Appropriateness scores for certain anonymization parameters in two of the three 
real-life event logs. Best results were achieved using the \sepsis log, while 
\coselog yielded partially better results. \bpic performed approximately equal 
for both distance measures. 

We see similar results when considering the Truly Sampled Score. Again, \sepsis 
shows the biggest improvements, for a range of \textit{k} and \textit{t}, 
whereas \coselog improved for selected parameter combinations. No noticeable 
differences can be observed for \bpic. Hence, regarding RQ1, we can say, that 
our embedding-based measure can preserve more control-flow utility. However, 
applying it gives no guarantee of achieving this aim.

\begin{wrapfigure}{r}{0.5\textwidth}
	\vspace{-1em}
	\begin{center}
		\includegraphics[width=\linewidth]{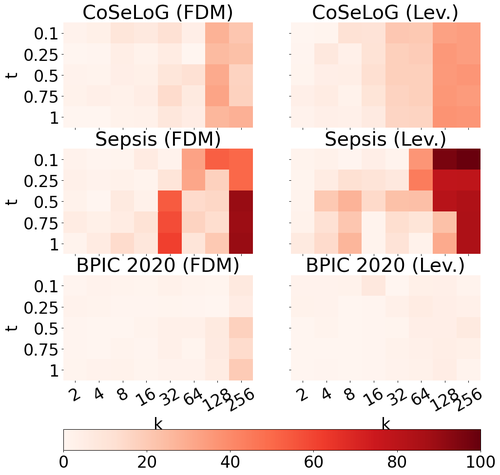}
	\end{center}
	\vspace{-1em}
	\caption{Total Duration Error of anonymized logs.}
	\label{fig:TDE}
	\vspace{-1em}
\end{wrapfigure}

Additionally, we want to investigate, if the application of the  
embedding-based measure harms other utility aspects of the log (RQ2). By 
considering the duration errors depicted in \autoref{fig:TDE}, 
we observe the following: For 
\sepsis, while results for specific scenarios vary widely, showing huge 
improvements in some cases and comparable declines in others, the global error neither increased, nor decreased.
Rather negligible improvements were made for \coselog, whereas 
both approaches are comparable for \bpic.
Concluding, in regards to RQ2, our results do 
not provide evidence that applying a distance measure focused on control-flow 
aspects will have negative implications for other utility aspects of the event 
log.

Finally, to address RQ3, we take a closer look at the internal properties of our 
distance measure. The embedding-based distance measure captures the 
context of each event in a feature vector. Similar feature vectors of two 
events indicate that the activities referenced in the events often appear in a 
similar context and thus, are semantically exchangeable to a certain extent. An 
event log with many similar events therefore 
enables the embedding-based distance measure to guide anonymization effectively 
in the selection of traces to merge, whereas event logs where events relate to 
activities of disjunct contexts, lack such possibility.

To investigate this, we first consider the average and standard deviations of the computed event distances per log, shown in \autoref{tab:internalproperties}. 
The table reveals that events in \bpic are particularly distant from each 
other, with the highest average (0.93)  and smallest standard deviation (0.28), 
whereas the \sepsis log has the lowest average and highest standard deviation. 
This reveals that for the latter, our approach using the embedding-based 
distance is able to have a greater impact on the anonymization procedure when 
compared to an approach based on the Levenshtein distance, which is indeed 
confirmed by the larger positive impact that our approach has for this log, 
when compared to the \bpic and \coselog logs.

\autoref{tab:internalproperties} furthermore shows the 
median share of the $x$ most frequent directly followers per activity (as a fraction of its total number of directly followers).
This way, we measure how similar the contexts
of activities are in a log. The activities of \bpic have the most common 
context, backing up the high distinctiveness of them. The \sepsis  log on the 
other hand, are on average noticeably closer in distance and share less 
neighbouring events. Allowing the measure to distinguish stronger between the 
closeness of activities. Consequently, we saw higher utility preservation in 
the 
previous experiments for the \sepsis log, compared to the Levensthein 
distance-based baseline. With \coselog falling both in terms of utility 
preservation and internal measures between the two other logs.

\begin{table}[ht!]
	\centering
	\vspace{-2em}
		\caption{Internal properties of our measure in terms of the distance between events and the share of the top-$x$ most frequent followers per activity (median)}
	\label{tab:internalproperties}
	\begin{tabular}{l@{\hskip 1.5em}c@{\hskip 1.5em}c @{\hskip 1.5em}c@{\hskip 1.5em}c@{\hskip 1.5em}c}
		\toprule
		&\multicolumn{2}{l}{\textbf{Event distance}} &  \multicolumn{3}{c}{\textbf{Share of top-\textit{x} followers}}\\
		\textbf{Event log}& \textbf{Avg.} & \textbf{Stdev.} & 
		$\mathbf{x=1}$ & $\mathbf{x=2}$ &$\mathbf{x=3}$ \\		
		\midrule
		\coselog & 0.83 & 0.29 & 82.8\% & 91.3\% & 99.2\% \\
		\sepsis & 0.77 & 0.42  & 67.0\% & 82.1\% & 94.3\% \\
		\bpic & 0.93 & 0.28 & 88.8\% & 96.7\% & 99.0\% \\
		\bottomrule
	\end{tabular}
	\vspace{-2em}
\end{table}

\section{Related Work}
\label{sec:related_work}

Privacy-preserving process mining received much attention 
recently~\cite{elkoumy2021privacy}. In particular, the problem of anonymizing 
event logs was addressed in several approaches, since logs generally show 
serious re-identification risks~\cite{von2020quantifying}. PRETSA~\cite{PRETSA} 
has been proposed as an algorithm to tackle this problem 
based on the privacy notions of $k$-anonymity and $t$-closeness. Similarly, 
Rafiei et al.~\cite{rafiei2020tlkc,rafiei2021group} 
proposed alternative techniques to ensure a group-based privacy guarantee, 
inspired by $k$-anonymity. A similar approach to group-based privacy protection 
is also realized in~\cite{batista2021uniformization}.
Either way, event log anonymization relies on a distance measure to decide on 
which traces to merge with each other and, so far, only the Levenshtein 
distance as a simple syntactic measure was considered. We addressed this 
shortcoming with the embedding-based measure presented in this work. 

Several studies also focused on differential privacy, an alternative privacy 
guarantee, that limits the impact one individual may have on the data. 
While~\cite{mannhardt2019privacy} and~\cite{kabierski2021} focus on specific 
queries performed on the event logs, the issue of publishing differential 
private event logs was covered by~\cite{pripel} and~\cite{elkoumy2021mine}. 

Instead of anonymizing event logs, privacy-preserving algorithms for 
process mining may be employed to protect the data of individual 
persons~\cite{fahrenkrog2019providing}. Here, the distributed analysis 
of the control-flow of event logs was explored in~\cite{elkoumy2020secure} 
and~\cite{liu2016towards}, whereas privacy-aware role mining was studied 
in~\cite{RafieiA19}.

The application of privacy-preserving process mining in the healthcare domain 
was studied in~\cite{pika2020privacy}, highlighting the importance of the 
ability to customize the algorithms to domain-specific requirements. To foster 
the uptake of privacy-preserving process mining, the respective techniques have 
been made accessible as tools, including ELPaaS~\cite{bauer2019elpaas}, 
Shareprom~\cite{elkoumy2020shareprom}, and the tool described 
in~\cite{RafieiA20a}.

\section{Conclusion}
\label{sec:conclusion}

In this paper, we presented an embedding-based trace distance measure that is 
tailored to event log anonymization. It leverages feature vectors learned by 
the Act2Vec model to assess the similarity of events and, hence, traces. Unlike 
syntactic distances commonly used in event log 
anonymization, it thereby incorporates the context in which events occur. 
Moreover, the measure considers differences in the trace lengths in an 
asymmetric manner to 
guide the selection of which trace to merge into which other trace as part of 
the anonymization. 
Our experimental results indicate that an embedding-based distance 
measure can indeed improve the results of event log anonymization, compared to 
the use of the Levenshtein distance.
Specifically, the improvement is most pronounced for event logs that contain 
events of different activities that frequently appear in similar contexts.

\bibliography{references}
\bibliographystyle{splncs04}
\end{document}